\documentclass[aps,showpacs,superscriptaddress]{revtex4}
\usepackage{graphicx}

\begin{document}

\newcommand{\be}   {\begin{equation}}
\newcommand{\ee}   {\end{equation}}
\newcommand{\ba}   {\begin{eqnarray}}
\newcommand{\ea}   {\end{eqnarray}}
\newcommand{\tr}   {{\rm tr}}

\title{Pure State Correlations: Chords in Phase Space}

\author{Alfredo M. Ozorio de Almeida}
\email{ozorio@cbpf.br}
\homepage{www.cbpf.br/~ozorio}

\author{Ra\'ul O. Vallejos}
\email{vallejos@cbpf.br}
\homepage{www.cbpf.br/~vallejos}
\affiliation{ Centro Brasileiro de Pesquisas F\'{\i}sicas (CBPF), \\
              Rua Dr.~Xavier Sigaud 150, 22290-180 Rio de Janeiro, Brazil}

\author{Marcos Saraceno}
\email{saraceno@tandar.cnea.gov.ar}
\homepage{www.tandar.cnea.gov.ar/grupos/QCG}
\affiliation{Departamento de F\'{\i}sica, 
             Comisi\'on Nacional de Energ\'{\i}a At\'omica, \\
             Avenida del Libertador 8250, (1429) Buenos Aires, Argentina}
\date{\today}

\begin{abstract}
The intensity of the overlap of a quantum state with all its phase
space translations defines its quantum correlations.
In the case of pure states, these are invariant with respect to 
Fourier transformation.
The overlaps themselves are here studied in terms of the Wigner
function and its Fourier transform, i.e., the characteristic function
or chord function.
Unlike the Wigner function, the chord function need not be real,
but eventual symmetry with respect to reflections about a phase
space point may relate these representations.
Semiclassical approximations for the ``classical-like" region of 
small chords and for large chords are derived.
These lead to an interpretation of the Fourier invariance in terms
of conjugate chords.
The interrelation of large and small (sub-Planck) phase space structures
previously noted in the literature are thus reinterpreted.
\end{abstract}



\maketitle

\section{Introduction}
\label{section1}

Phase space correlations in quantum states have quite different
properties than correlations in classical distributions functions. 
At the quantum level, the uncertainty principle imposes, on one hand, 
limitations to the possible distribution functions and at the same 
time creates strong relations between small and large scales that 
result in peculiar properties.
Thus, for example, it has been realized \cite{zurek01} that the phase 
space area $A$ over which a pure state extends determines the minimal 
size of the high frequency oscillation structures $\delta A$ in the 
Wigner function by a kind of complementarity relationship, 
\begin{equation}
\label{eq1.1}
A\cdot \delta A \gtrsim (2\pi \hbar )^2 
\end{equation}
(for one degree of freedom), thus characterizing these structures 
as ``sub-Planck''.
This relationship between possibly macroscopic areas $A$ and 
sub-Planck areas $\delta A$ is entirely due to the finiteness 
of Planck's quantum of action and is not present when considering 
the correlations of classical distribution functions. 
In general, and especially for systems with more than one degree
of freedom, we shall show that it is more revealing to relate a
given large displacement to a specific small scale oscillation,
rather than to relate areas.

This complementarity between small and large scales in quantum
distribution functions can be accessed by the Wigner function and its
Fourier transform. Moreover, the latter, sometimes referred to as the
characteristic function, the generating function, or simply the chord
function, also lies at the core of a full represention of quantum
mechanics, on a par with the Weyl-Wigner representation itself. 
Further considerations about the corresponding conjugate 
classical phase spaces as well as the notation is found in 
Appendix \ref{appA}. 
In this paper we compare the way that both these representations
exhibit quantum coherences on all scales, that are overlaid on purely 
classical structures. 
The copious previous litterature on the Wigner functions allows us to 
concentrate on the properties of the chord representation.

In Section \ref{section2} we present the intrinsic definition of phase 
space correlations for a general density operator $\hat{\rho}$ to be 
studied.
These are then related to the chord function and the Wigner function
in Section \ref{section3}. Then in Section \ref{section4}, we exploit 
the general interrelation of these functions in the case of pure states 
and present some simple examples. 
Section \ref{section5} is dedicated to the theory of small chords. 
This is the classical region, corresponding to a caustic of the Wigner
function. 
The simple approximation thus obtained is rederived in the next section 
within the full semiclassical theory valid for large chords.
Squaring the semiclassical chord function in section \ref{section7}, 
we interpret the invariance of the correlations of a pure state with 
respect to the Fourier transform:
This results from a geometrical conjugacy of chords defined on the 
classical phase space structures. 
Thus the complementarity of large and small chords arises
as a limit among  finite  conjugate chords. 
The discussion in the last section 
recapitulates the full picture for extended pure states including 
``ergodic states" of chaotic Hamiltonians.

\section{Phase Space Correlations}
\label{section2}

The correlation between two states of a quantum system, represented
by their density operators, $\hat{\rho}_A$ and $\hat{\rho}_B$, can be 
given an intrinsic definition as 
\begin{equation}
\label{eq2.1}
C_{AB}\equiv \frac{\tr \hat{\rho}_A   \,     \hat{\rho}_B}
            {\sqrt{\tr \hat{\rho}_A^2 \, \tr \hat{\rho}_B^2}} \;.
\end{equation}
By the Schwartz inequality this quantity is always less than unity
and for pure states, 
$\hat{\rho}=| \psi \rangle \langle  \psi |$,  
it reduces  to 
$C_{AB} = | \langle  \psi_A|\psi_B \rangle|^2$. 
When the pair of states are obtained by the unitary evolution of 
slightly different hamiltonians from the same initial state, 
the behaviour in time of $C_{AB}$ reflects the regular or 
chaotic nature of the underlying classical motion 
\cite{peres84, loschmidt, garciamata04}.
If the states are related by a unitary transformation generated by
an Hermitian operator $\hat K$, 
\begin{equation}
\label{eq2.2}
\hat{\rho}_\alpha = e^{-i \alpha \hat{K} / \hbar}
                    \hat{\rho}  
                    e^{ i \alpha \hat{K} / \hbar}  \; ,
\end{equation}
with $\alpha$ a real parameter, the small--$\alpha$ behaviour 
of the correlation is easy to ascertain: 
\begin{equation}
\label{eq2.3}
C(\alpha ) \equiv
\frac{                              \tr  \hat{\rho} \,
       e^{-i \alpha \hat{K} / \hbar} \hat{\rho} \,
       e^{ i \alpha \hat{K} / \hbar}
     }{  \tr  \hat{\rho}^2} 
\approx 1 - \frac{\alpha^2}{\hbar^2} 
            \frac{ \tr [\hat{\rho}^2 \hat{K}^2-
                     \hat{\rho}\hat{K}\hat{\rho}\hat{K}]}{ \tr \hat{\rho}^2}
=       1 + \frac{\alpha^2}{2\hbar^2} 
            \frac{ \tr [\hat{\rho},\hat{K}]^2}{ \tr \hat{\rho}^2} \le 1 \; .
\end{equation}
If $\hat \rho$ represents a pure state, this quadratic behaviour 
relates $C(\alpha)$ to the dispersion of the generator $\hat K$,
i.e., 
$\langle \hat{K}^2 \rangle - \langle \hat{K} \rangle^2$
\cite{alonso04}. 

In what follows we will be concerned with the phase space
correlations of a quantum state produced by unitary translations
in the $L$--dimensional phase space
$(p,q)=(p_1,\ldots,p_L,q_1,\ldots,q_L)$. 
Denoting the corresponding quantum operators by 
$(\hat{p},\hat{q})$, the translation operators read
\begin{equation}
\label{eq2.4}
\hat{T}_{\xi} =   \exp \left[ \frac{i}{\hbar}
                \left( \xi_p \cdot \hat{q} -
                       \xi_q \cdot \hat{p}
               \right)\right]\ ,
\end{equation}
where the vector, or chord, $\xi=(\xi_p,\xi_q)$ represents an arbitrary 
direction in phase space.
The expression
\begin{equation}
\label{eq2.5}
C_{\xi} \equiv  \frac{   \tr  \hat{\rho} T_{\xi} \hat{\rho} T_{\xi}^\dagger}
                     {   \tr  \hat{\rho}^2} 
        = \frac{   \tr  \hat{\rho}  \hat{\rho}_\xi}{  \tr  \hat{\rho}^2} 
\end{equation}
is an intrinsic definition of phase space correlation, quite independent 
of the representation used to compute it and should not be confused with,
e.g., {\em local} wave function correlations \cite{wfc}, which do depend 
on the specific coordinate representation. 

In quantum optics it is customary to switch to the basis of creation and
anihilation operators 
$(\hat{q}\pm i\hat{p})/\sqrt{2 \hbar}$. 
In this context, the translation operator (\ref{eq2.4}) depends on the 
complex chords $(\xi_p \pm i \xi_q)/\sqrt{2 \hbar}$ and is called the 
displacement operator \cite{schleich, glauber63}.
The semiclassical limit for a complex phase space is not as transparent 
as the real theory treated here. 
However it is quite feasible to effect phase space translations in the
optical context \cite{lutterbach97}.

Some of the general formulae in the following sections are related to the
theory of gaussian noise channels. Indeed, the ``fidelity" of a gaussian channel
is just the correlation (\ref{eq2.5}) averaged with a gaussian window for
the chords $\xi$ \cite{caves04}.   
%

\section{The Chord Function and the Wigner Function}
\label{section3}

The chord symbol for an operator $\hat{A}$ is defined as
\begin{equation}
\label{eq3.1}
A(\xi) \equiv   \tr  \hat{T}_{-\xi}\hat{A}\ ,
\end{equation}
allowing for the complete representation of $\hat A$ in terms 
of the unitary translations in phase space:
\begin{equation}
\label{eq3.2}
\hat{A} = \int \frac{d\xi}{(2\pi\hbar)^L} A(\xi) \hat{T}_{\xi} \; .
\end{equation}
In the case of the density  operator $\hat{\rho}$, it is convenient 
to alter the normalization, so that 
\begin{equation}
\label{eq3.3}
\chi(\xi )\equiv  \frac{1}{(2\pi\hbar)^L} \tr  \hat{T}_{-\xi}\hat{\rho}
\end{equation}
is the definition of the chord function, also known as the
characteristic function, or the generating function. (In classical
mechanics the displacement $\xi$ results from a trajectory of
which it is the chord -- see Appendix \ref{appA}.) 

The Fourier transform of the chord symbol is the Weyl symbol $A(X)$,
\begin{equation}
\label{eq3.4}
A(X)= \int \frac{d\xi}{(2\pi \hbar )^L} \ 
      e^{-i X \wedge \xi / \hbar} A(\xi ) \; ,
\end{equation}
in terms of the skew product
\begin{equation}
\label{eq3.5}
X \wedge \xi = P \cdot \xi_q - Q \cdot \xi_{p} \; .
\end{equation}
In the case of the density operator, the Fourier 
transform of (\ref{eq3.3}) is the familiar Wigner function. 
But, since the Fourier transform of the translation operator 
$\hat{T}_{\xi}$ itself corresponds to the classical reflection 
through the phase space point $X$ \cite{ozorio98}, 
\begin{equation}
\label{eq3.6}
\int \frac{d\xi}{(2\pi \hbar )^L} \ 
          e^{i X \wedge \xi / \hbar} \ \hat{T}_\xi =
2^L \hat{R}_X  \; ,
\end{equation}
it follows that \cite{royer77}
\begin{equation}
\label{eq3.7}
W(X) \equiv \frac{1}{(\pi \hbar)^L} \tr  \hat{R}_X \hat{\rho} \; .
\end{equation}

Though translations and reflections are quite distinct operators,
they combine to form the affine group in geometry \cite{coxeter}, 
which is transported into quantum mechanics by the operators 
$\hat{R}_X$ and $\hat{T}_{\xi}$ \cite{ozorio98}. 
The family resemblance is striking when viewed, for instance, 
from the position representation: 
\begin{equation}
\label{eq3.8}
2^L \hat{R}_X = \int d \xi_q 
      \left| Q + \frac{\xi_q}{2} \right\rangle 
\left\langle Q - \frac{\xi_q}{2} \right|
       e^{i P \cdot \xi_q / \hbar}  \; ,
\end{equation}
whereas
\begin{equation}
\label{eq3.9}
\hat{T}_{\xi}= \int dQ \left| Q + \frac{\xi_q}{2} 
\right\rangle \left\langle    Q - \frac{\xi_q}{2} \right|
                e^{i \xi_p \cdot Q/ \hbar}  \; .
\end{equation}

It is well known that the Wigner function cannot be indentified with
a probability distribution in phase space, even though this
interpretation holds for marginal distributions and the calculation 
of averages of observables as phase space integrals. 
The main problem is that $W(X)$ can assume negative values and indeed 
they are present for all pure states that are not gaussian (coherent) 
states. 
For this reason, it is also improper to refer to $\chi(\xi)$ as a 
charactersitic function, even though its derivatives do generate 
moments of $\hat{q}$, $\hat{p}$, and any polynomial function in phase 
space, taking proper care of the operator ordering; for instance, 
\begin{equation}
\label{eq3.10}
\langle  q^n \rangle \ =  \tr \hat{q}^n \hat{\rho} = 
(i\hbar )^n \ \frac{\partial^n}{\partial \xi_p^n} 
\ \chi(\xi) \Big|_{\xi = 0} \; .
\end{equation}
For the  case of the identity operator, we obtain the normalization 
condition: 
\begin{equation}
\label{eq3.11}
1 = \tr \hat{\rho} = \int dX \ W(X) = (2 \pi \hbar)^L \chi(0) \; .
\end{equation}
In the case of operators representing observables with smooth, 
classical-like Wigner functions, e.g., polynomials in $q$ and $p$, 
their corresponding Fourier transforms, i.e., the chord symbols
are sharply localized (improper functions) close to the origin.
On the other hand, the chord function for a normalized state is 
a proper function, and extends away from the origin, so as to represent 
truly quantum correlations.
The process of decoherence, destroying the purity of the initial 
state, generically washes away the exterior structure of the chord
function and compacts the mixed state onto the classical origin of 
chords. 
This has been shown for linear Markovian systems \cite{brodier04}, 
and will be the subject of further work. 
Here we will describe the large and small scale features of the chord 
function and their close intertwining.

Combining the definition (\ref{eq3.6}) of the Wigner function with 
the group properties of the translation and reflection operators 
\cite{ozorio98} it is easy to see that the Wigner function corresponding to 
the translated state
$\hat{\rho}_{\eta}=\hat{T}_{\eta}\hat{\rho}\hat{T}_{-\eta}$ 
is 
\begin{equation}
\label{eq3.12}
W_{\eta}(X)=W(X-\eta ) \; ,
\end{equation}
whereas the corresponding chord function is just
\begin{equation}
\label{eq3.13}
\chi_{\eta}(\xi ) = e^{i \eta \wedge \xi / \hbar} \chi(\xi ) \; .
\end{equation}

Unlike the Wigner function, the chord function is not necessarily
real, but it may be real for a particular choice of phase space
origin. It is shown in Appendix \ref{appB} that the necessary and 
sufficient condition for this is that there exists a symmetry 
centre $X$, such that $[\hat{\rho},\hat{R}_X]=0$, and it is chosen 
as the origin. 
Since $R_X$ has eigenvalues $\pm 1$, the pure or mixed state
$\hat{\rho}$, must then lie in the Hilbert subspace of either even 
or odd parity. 
For these parity-symmetric states, $\hat{\rho}_\pm$, the Wigner
function and the chord function are obtained from each other by a
mere rescaling:
\begin{equation}
\label{eq3.14}
W_{\pm}(X) = \pm \ 2^L\chi_{\pm}(-2X) \; .
\end{equation}  
For general unsymmetric states, the real part of the chord
functions is still determined by the diagonal part of $\hat{\rho}$ 
with respect to parity, whereas the imaginary part depends on the
off-diagonal part. 
However, it is the intensity of the chord function, $|\chi(\xi )|^2$, 
that turns out to be most useful.

The intrinsic definition of phase space correlations $C_{\xi}$ 
in (\ref{eq2.5}) is readily translated into the properties of Wigner
functions and chord functions: 
\begin{equation}
\label{eq3.15}
 \tr \; \hat{\rho} \; \hat{T_\xi} \; \hat{\rho} \; \hat{T}^\dagger_\xi=
(2\pi \hbar )^L \int dX \; W(X) \; W(X-\xi)=
(2\pi \hbar )^L \int d\eta \; e^{ i \eta \wedge \xi / \hbar}
\left| \chi(\eta )\right|^2 \; .
\end{equation}
Thus the correlations of the Wigner function can be identified with 
$C_{\xi}$ and $|\chi(\eta )|^2$ is just the power spectrum of $W(x)$.

\section{Pure States}
\label{section4}

It is worthwhile to recollect some examples for which the chord
function can be identified with the Wigner function, once the origin
is translated to the symmetry center. 
In all the following cases we consider states of a harmonic oscillator 
with one degree of freedom and unit mass.

i) Coherent states, $|\eta \rangle$, are displacements of the ground
state of the harmonic oscillator by $\hat{T}_{\eta}$. The Wigner
function is just a gaussian centered on $\eta$,
\begin{equation}
\label{eq4.1}
W_{\eta}(X) =
 \frac{1}{\pi \hbar}
 \exp \left[
-\frac{\omega}{\hbar}    \ \left(Q-\eta_q\right)^2 -
 \frac{1}{\hbar \omega}  \ \left(P-\eta_p\right)^2      
     \right] 
 \stackrel{\omega=1}{\longrightarrow}
 \frac{1}{\pi \hbar} e^{-(X-\eta)^2/\hbar} \; ,
\end{equation}
whereas, using (\ref{eq3.12}) and (\ref{eq3.13}),
\begin{equation}
\label{eq4.2}
\chi_{\eta}(\xi )= 
\frac{1}{2\pi \hbar}
 \exp 
 \left(\frac{i \eta \wedge \xi}{\hbar}\right)
 \exp \left[-\frac{\omega}{\hbar}
 \left(\frac{\xi_q}{2}\right)^2-\frac{1}{\hbar\omega}
 \left(\frac{\xi_p}{2}\right)^2\right] 
\stackrel{\omega=1}{\longrightarrow}
\frac{1}{2\pi \hbar}
e^{i \eta \wedge \xi / \hbar}
e^{-\xi^2 / 4\hbar} \; .
\end{equation}
So, any translation of the coherent state merely alters 
the phase of the gausssian chord function that sits on the origin.

(ii) A superposition of a pair of coherent states, 
$|\eta \rangle \pm |- \eta \rangle$ 
is sometimes known as a ``Schr\"{o}dinger cat state". 
Its Wigner function is (here and below we set $\omega =1$)
\begin{equation}
\label{eq4.3}
W_{\pm}(X) = 
\frac{1}{2\pi \hbar \,(1\pm e^{-\eta^2/\hbar})}
\left[e^{-(X-\eta )^2 / \hbar } +
      e^{-(X+\eta )^2 / \hbar } \pm 
    2 e^{-X^2         / \hbar } \cos \frac{2}{\hbar} X\wedge \eta \right] \; .
\end{equation}
It consists of two ``classical" gaussians centred on $\pm \eta$ and 
an interference pattern with a gaussian envelope centred on their 
midpoint.
The frequency of this oscillation increases with the separation 
$|2 \eta |$. 
For the chord function,
\begin{equation}
\label{eq4.4}
\chi_{\pm}(\xi ) = 
\frac{1}{4\pi \hbar \,(1\pm e^{-\eta^2/\hbar})}
\left[e^{-(\xi/2-\eta )^2 / \hbar } +
      e^{-(\xi/2+\eta )^2 / \hbar } \pm 
    2 e^{-\xi^2       / 4\hbar } 
\cos \frac{1}{\hbar} \xi \wedge \eta \right] \; ,
\end{equation}
this same configuration has to be reinterpreted. 
Now the internal correlations of the individual coherent states are
both superimposed onto the neighbourhood of the origin, as in (i), 
while their cross-correlation generates new gaussians centred on 
the separation vectors $\pm 2\eta$. 
Of course, the general case of coherent states 
$|\eta_1 \rangle$ and $|\eta_2\rangle$ 
merely leads to gaussians centred on
$\pm (\eta_1-\eta_2)$ with addition of the phase factor 
$\exp [i (\eta_1+\eta_2) \wedge \xi  /2\hbar]$. 

(iii) 
Fock states, $|n \rangle$, i.e., the excited states of 
the harmonic oscillator, also have reflection symmetry with respect 
to the origin. 
Thus, from the exact Wigner function, first derived by
Gr\"onewold \cite{gronewold46},
\begin{equation}
\label{eq4.5}
W_n(X) = 
\frac{(-1)^n}{\pi \hbar} e^{-X^2/\hbar}
L_n \left(\frac{2X^2}{\hbar}\right) \; ,
\end{equation}
where $L_n$ is a Laguerre polynomial, and (\ref{eq3.14}) we obtain 
the chord function
\begin{equation}
\label{eq4.6}
\chi_n (\xi ) = \frac{e^{-\xi^2/4\hbar }}{2\pi \hbar}
L_n \left( \frac{\xi^2}{2\hbar} \right) \; .
\end{equation}
It is interesting to note that the symmetry centre, which produces
the maximum amplitude of the Wigner function is nowhere near the
classical  manifold with energy 
$E_n=\left(n+\frac{1}{2}\right) \hbar\omega$. 
However, this point lies in a region of narrow oscillations, 
so that it does not affect the average of smooth observables.

All the above examples are singled out by some point of reflection
symmetry, which must always be chosen as the origin for the chord
function to be real. The chord function always assumes its maximum
value $1/(2\pi \hbar)^L$ at the origin, whatever the symmetry.
For a pure state the proof is immediate because
\begin{equation}
\label{eq4.7}
\langle \psi |\psi_\xi \rangle\ = (2\pi \hbar )^L \chi(-\xi )\ ,
\end{equation}
which cannot have modulus greater than one, whatever the symmetry. 
But even an average of overlaps cannot exceed one, so $\chi(0)$ is
also the maximum for mixed states.
The Wigner intensity $[W(X)]^2$, need not have such a prominent peak 
in general. 
However we shall see in the section \ref{section6} that the large scale 
features of the semiclassical forms of the Wigner function and the 
chord function maintain a mutual correspondence, even in the absence 
of a reflection symmetry.

General invariance with respect to Fourier transformation does hold
for the correlation in the case of pure states. 
Indeed, combining (\ref{eq4.7}) with (\ref{eq3.15}) we obtain
\begin{equation}
\label{eq4.8}
(2\pi\hbar)^{2L} C_\xi=
|\chi(\xi )|^2=\int \frac{ d \eta}{(2\pi\hbar)^L} \;
e^{i \eta \wedge \xi / \hbar}
|\chi(\eta )|^2 \; .
\end{equation}
This is a remarkable property of all pure states and is in no way
restricted by special symmetry properties that relate certain Wigner
functions to their respective chord functions.
All the same, we can start by considering the example of the chord 
functions so far studied:
(i) in the case of a single gaussian (\ref{eq4.1})
the invariance is obvious, because the square modulus is 
a gaussian with the appropriate width for its Fourier 
transform to be of the same form.
(ii) For a pair of coherent states, the chord function is a 
sum of gaussians. Its square modulus is also gaussian and we
again return to the same function by Fourier tranformation.
(iii) For Fock states this is not so obvious, but we can also write 
$|\chi(\xi )|^2$ as 
\begin{equation}
\label{eq4.9}
|\chi(\xi )|^2 = 
\sum_k  a_k 
\frac{\partial^k}{\partial \lambda^k} 
e^{-\lambda \xi^2/2\hbar} \,  \bigg|_{\lambda=1}\; ,
\end{equation}
with coefficients $a_k$ that do not depend on $\xi$.
So, the invariance of the ground state gaussian
entails that of all Fock states. 

The Fourier invariance condition (\ref{eq4.8}) includes as a special case 
the more familiar tracing over the full pure state condition 
$\hat{\rho}^2=\hat{\rho}$. 
It can be easily checked that setting $\xi=0$ in (\ref{eq4.8}) 
gives the chord representation of the identity
$\tr \hat{\rho}^2= \tr \rho$.
It follows that the difference of both sides of (\ref{eq4.8}) for 
each chord $\xi$ is a measure of the purity that generalizes the 
linear entropy. 
However, the loss of the phase information contained in the chord 
function, but absent in $|\chi(\xi)|$, implies that these are necessary 
conditions, whereas the full sufficient condition of purity is only 
$\hat{\rho}^2=\hat{\rho}$, 
which is expressed in the chord representation as 
\begin{equation}
\label{eq4.10}
\int d\eta \, \chi(\eta )\, \chi(\xi -\eta ) \,
e^{i \xi \wedge \eta /2\hbar} = 
\int d\eta \, \chi_{\xi/2}(\eta ) \, \chi(\xi -\eta ) = 
\chi(\xi ) \; ,
\end{equation}
with $\chi_{\xi/2}(\eta )$ defined by (\ref{eq3.13}).

\section{Small Chords}
\label{section5}

We now consider the chord function and the correlation function
of semiclassical states, i.e., those related to generalized
Bohr-Sommerfeld energy levels. 
This section focuses on the limit of small chords.

Starting from the Wigner-Weyl representation we may rewrite
\begin{equation}
\label{eq5.1}
\chi(\xi )=
\frac{1}{(2\pi\hbar)^L} \; \langle \hat{T}_{-\xi}\rangle =
\frac{1}{(2\pi\hbar)^L}    \int dX \ T_{-\xi}(X)W(X)\ ,
\end{equation}
Besides the Wigner function, we have here introduced the Weyl symbol
for the translation operator: 
\begin{equation}
\label{eq5.2}
2^L \, \tr \, \hat{T}_{-\xi} \, \hat{R}_X \equiv 
T_{-\xi}(X) =
e^{-i \xi \wedge X / \hbar } \; .
\end{equation}
If $\xi$ is small enough, i.e., $|\xi| \lesssim \hbar$, then 
$T_\xi(X)$ behaves like a smooth, classical-like symbol. 
It still oscillates, but with a classical wavelength. 
In such a case, to a good approximation, we can replace the 
Wigner function in (\ref{eq5.1}) by the simplest semiclassical 
expression \cite{berry77}
\begin{equation}
\label{eq5.3}
W_\mathcal{I}(X) \simeq 
\frac{1}{(2\pi )^L}\ \delta \left( I(X)-\mathcal{I} \right ) \; .
\end{equation}
Here $I(X)$ is the set of $L$ action variables for an integrable
system with $L$ degrees of freedom and $\mathcal{I}$ is the set of 
quantized action values for this particular state 
\cite{berry77, ozorio}. 
Within this approximation, the average of a quantum observable 
$\hat A$ is just a purely classical average over a torus:
\begin{equation}
\label{eq5.4}
\langle  \hat{A} \rangle  \simeq 
\int dX \ A(X) \ \frac{\delta \left( I(X)-\mathcal{I} \right)}{(2\pi )^L}=
\int \frac{d \theta}{(2\pi )^L} \ A(\theta)  \; ,
\end{equation}
where $\theta$ are the angle variables conjugate to the actions, which
describe positions on the quantized torus, and $A(\theta)=A(X(\theta))$. 
In the case of the representation for the chord function (\ref{eq5.1}), 
we obtain
\begin{equation}
\label{eq5.5}
\chi(\xi ) \ \simeq 
\frac{1}{(2\pi\hbar)^L} 
\int \frac{ d\theta}{(2\pi )^L}
\ e^{-i \xi \wedge X(\theta )/\hbar } \; .
\end{equation}
Certainly, a bad choice of origin will lead to large phases in
(\ref{eq5.5}), but we have already studied the trivial phase change due
to translating the origin. 
Thus, we can increase the quality of 
(\ref{eq5.5}) by choosing the origin to minimize $|X(\theta)|$
on average.

Let us check the semiclassical approximation (\ref{eq5.5}) for the
simplest case of the Fock states, discussed in the previous section.
Choosing $\omega =1$, we have action-angle variables that are merely 
canonical polar coordinates 
\begin{equation}
\label{eq5.6}
q = \sqrt{2 \mathcal{I}} \ \cos \theta \; , \qquad 
p = \sqrt{2 \mathcal{I}} \ \sin \theta \; .
\end{equation}
Thus, choosing $\xi$ along the $p$-axis, without loss of
generality, we obtain
\begin{equation}
\label{eq5.7}
\chi(\xi) \simeq 
\frac{1}{2\pi\hbar} 
\int \frac{d\theta}{2\pi }
\ e^{- i \sqrt{2 \mathcal{I}} \, |\xi | \cos \theta / \hbar } =
\frac{1}{2\pi\hbar} 
J_0 \left( \frac{\sqrt{2 \mathcal{I}}\ |\xi |}{\hbar}\right) \; ,
\end{equation}
where $J_0$ is a Bessel function.
This result should be a good approximation to the exact formula 
(\ref{eq4.6}) for $|\xi| \lesssim \hbar$ and large 
$n=\mathcal{I}/\hbar-1/2$.
To check this, first note that $\chi(\xi)$ oscillates with
a wavelength $\lambda \sim \hbar / \sqrt{\mathcal{I}}$. 
In terms of this scale, and defining $z=|\xi|/\lambda$, the 
argument of the Laguerre polynomial in (\ref{eq4.6}) reads 
\be
\frac{\xi^2}{2\hbar}=\frac{z^2}{2n} \; .
\ee
Hence, for $z^2 \ll n$, we recover (\ref{eq5.7}) by
using the formula 
\cite{abramowicz}
\be
\lim_{n \to \infty} L_n\left(\frac{z^2}{2n}\right) = 
J_0\left(\sqrt{2}z\right) \; .
\ee

In the next section we shall study the semiclassical limit of the 
chord function for large chords and show that in the particular case 
of the harmonic oscillator this coincides with the expansion of 
(\ref{eq5.7}) for large argument:
\begin{equation}
\label{eq5.8}
J_0 (y) \approx 
\frac{2}{\sqrt{\pi y}} \,  \cos \left(y-\frac{\pi}{4}\right) \; .
\end{equation}
%

\section{Semiclassical Theory}
\label{section6}

The easiest path to obtain the semiclassical form of the chord function
for finite chords is to start from the general WKB wave functions. 
The underlying classical structure is assumed to be a
curve (for $L=1$) or a Lagrangian surface ($L>1$) defined by the
actions $I$. 
Fixing the action, 
such structure can be described locally as a function $p=p(q,I))$ 
which may have several branches.
Thus we define 
\begin{equation}
\label{eq6.1}
S(q,I)= \int^q_{q_0}\ p(q^\prime,I) \, dq^{\prime}\ ,
\end{equation}
which is the generating function for the transformation 
$(p,q)\rightarrow (I,\theta )$, such that 
\begin{equation}
\label{eq6.2}
\frac{\partial S}{\partial I}= \theta \, , \qquad 
\frac{\partial S}{\partial q}= p \; .
\end{equation}
Then the corresponding WKB wave functions are linear
combinations of
\begin{equation}
\label{eq6.3}
\langle q| \psi_I \rangle \, = c
\left| \det \frac{ \partial^2 S(q,I)} 
                    {\partial q \partial I} \right|^{1/2} \, 
e^{i S(q,I)/\hbar } \; ,
\end{equation}
for the various branches of $S(q,I)$, with $c$ a normalization 
constant \cite{vanvleck28, ozorio}. 

The similarity between equations (\ref{eq3.8}) and (\ref{eq3.9}) for 
the operators that determine the chord function and the Wigner 
function, allows us to follow the same steps as Berry \cite{berry77}
for the semiclassical Wigner function. 
Thus we presume that
\begin{eqnarray}
\label{eq6.4}
\chi(\xi )& = &\frac{c^2}{(2\pi \hbar)^L} 
\int dQ^\prime \exp 
\left\{ \frac{1}{\hbar} 
\left[ S \left(Q^\prime + \frac{\xi_q}{2},I \right)-
       S \left(Q^\prime - \frac{\xi_q}{2},I \right)- \xi_p Q^\prime
\right] \right\} \nonumber \\
&& \left| \det \, \frac{ \partial^2 S(Q^\prime + \xi_q / 2,I) }
                       { \partial q \partial I}\ 
          \det \, \frac{ \partial^2 S(Q^\prime - \xi_q / 2,I) }
                       { \partial q \partial I}\ 
  \right|^{1/2}
\end{eqnarray}
has a stationary point within a single of branch of $S(q,I)$.
The stationary phase condition for $Q^\prime = Q$
\begin{equation}
\label{eq6.5}
p_\mathcal{I} \left( Q + \frac{\xi_q}{2}\right) - 
p_\mathcal{I} \left( Q - \frac{\xi_q}{2}\right) = \xi_p
\end{equation}
specifies that the geometrical chord corresponding to the arc of 
$\, p_\mathcal{I}(q)=p(q, I=\mathcal{I})$, 
lying between 
$q=Q-\xi_q/2$ 
and 
$q=Q+\xi_q/2$, 
coincides with the given chord $\xi$.

\begin{figure}
\includegraphics[width=17cm]{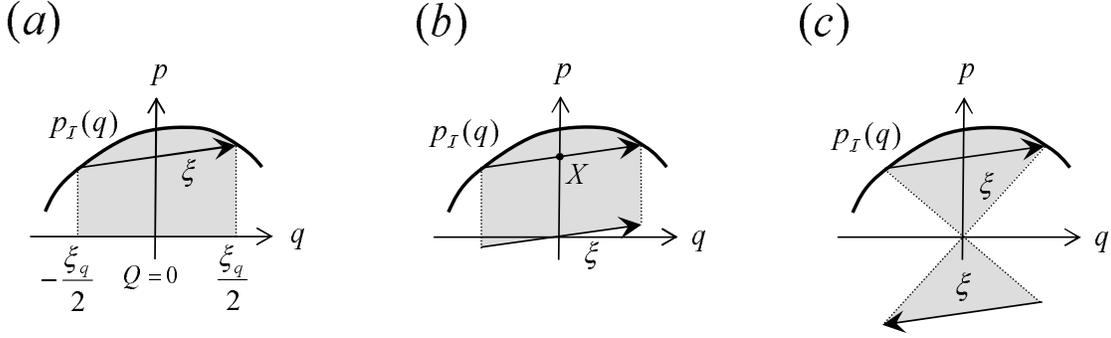}
\caption{The stationary phase condition for the semiclassical chord 
function $\chi(\xi)$ is just that $\xi$ be a geometrical chord for
the curve $I(x)=\mathcal{I}$, locally given by $p_\mathcal{I}(q)$.
The stationary phase itself is given by the shaded area in (a).
The constuctions in (b) and (c) are invariant with respect to
linear transformations and have the same areas.}
\label{fig1}
\end{figure}

Unlike the Wigner function, the phase of the chord function depends
on the choice of phase space origin, but this phase is
trivially specified by the overall phase factor (\ref{eq3.13}) for a
translation by a vector $\eta$. 
Thus, let us for now assume that the origin is translated to 
$(0,Q)$. 
Then the stationary phase action is just the area shown in 
Fig.~\ref{fig1}(a).
The alternative constructions (b) and (c), also shown in Fig.~\ref{fig1},
have the same area and enjoy the advantage that they survive arbitrary 
linear canonical transformations which preserve the origin. 
In other words, we may always consider the phase for the semiclassical 
chord functions to be defined by a sum of areas: 
(i) the area sandwiched between the chord and the arc into which if fits,
$S(X,\mathcal{I})$. 
This is just the same as the one for the semiclassical Wigner function, 
except that there the construction starts from the centre of the chord 
$X$ [Fig.~\ref{fig1}(b)], rather than from the chord itself.
(ii) The area that is added to this may be taken as that of the
parallelogram obtained by tranporting the centre of $\xi$ to the origin 
[Fig.~\ref{fig1}(b)], or the polygonal figure of eight constructed by
the chord and its reflection around the origin [Fig.~\ref{fig1}(c)]. 
In either case the value of the area is just $X \wedge \xi$. 
Therefore we construct the chord generating function
\begin{equation}
\label{eq6.6}
S(\xi ,I) = S(X,I) - X \wedge \xi \; ,
\end{equation}
which is just the Legendre transformation of the centre action,
appropriate to the Wigner function, as discussed in 
Appendix \ref{appA}.
This geometry is immediately generalized for $L>1$: the action
for any polygonal figure is just the algebraic sum of the areas
of its projections on each conjugate plane. 
Furthermore, it does not matter which arc is chosen between the
tips of $\xi$ along $p_\mathcal{I}(q)$ because this is a 
Lagrangian surface.  

Thus, the contribution of each realization of the chord $\xi$, 
with centres $X_j$ and closing arcs $\gamma_j$ on the quantized 
curve or Lagrange manifold $I(x)=\mathcal{I}$ has the form
\begin{equation}
\label{eq6.7}
\chi_j(\xi ) = 
A_j(\xi) \exp \left[
\frac{i}{\hbar} \, S_j(\xi )- \sigma_j \, \frac{\pi}{4}
             \right] \; .
\end{equation}
Here $\sigma_j$ is the signature of the matrix
\begin{equation}
\label{eq6.8}
\frac{\partial^2}{\partial Q^2}
\left[S(q_+,I)-S(q_-,I) \right]\ ,
\end{equation}
where $q_{\pm} = Q \pm \xi/2$, so that in the case of the curve in
Fig.~\ref{fig1} we have
$\sigma =+1$.

\begin{figure}
\includegraphics[height=10cm]{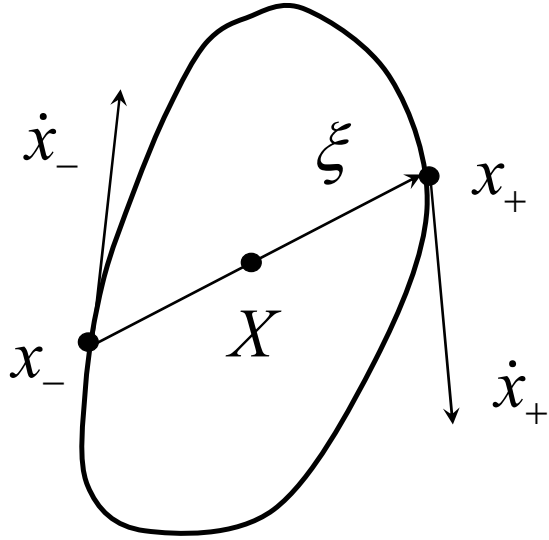}
\caption{The semiclassical chord function, $\chi(\xi)$, is determined
by the geometrical chord $\xi$, centered on $X$, and joining the points 
$x_-$ and $x_+$ on the quantized classical curve. 
The amplitude depends on both phase space velocities at the tips, as 
explained in the text.}
\label{fig2}
\end{figure}

The amplitude for each realization of the chord, within a
normalization factor, is given by
\begin{eqnarray}
\label{eq6.9}
\left| A_j(\xi) \right|^2 & = &
\left| \det \left[ \frac{\partial p_\mathcal{I}}{\partial q} (q_+)
                  -\frac{\partial p_\mathcal{I}}{\partial q} (q_-) 
           \right] \,
\det \frac{\partial I}{\partial p} \left( p(q_+),q_+) \right) \,
\det \frac{\partial I}{\partial p} \left( p(q_-),q_-) \right) \right|
\nonumber \\
& =& \left| \det \left[ \frac{\partial I}{\partial p}\Bigg|_+
                        \frac{\partial I}{\partial q}\Bigg|_- -
                        \frac{\partial I}{\partial p}\Bigg|_+
                        \frac{\partial I}{\partial q}\Bigg|_- 
                 \right] \right| \; ,
\end{eqnarray}
because %
\begin{equation}
\label{eq6.10}
\frac{dI}{dq} \left( p(q),q \right) =
\frac{\partial I}{\partial q} +
\frac{\partial I}{\partial p}\ 
\frac{\partial p}{\partial q} = 0 \; ,
\end{equation}
along the classical surface.
These amplitudes coincide with those of the Wigner function evaluated
at the point $X_j(\xi)$.
In the case where $L=1$, the $1\times 1$ determinant is only a
single factor. 
Taking $I(x)$ to be a Hamiltonian, the corresponding phase space 
velocity tangent to the phase space curve is
\begin{equation}
\label{eq6.11}
\dot{x}_I = J \, \frac{\partial I}{\partial x} \; ,
\end{equation}
where $J$ is the symplectic matrix (\ref{eqc.4}),
and we may interpret the amplitude of the chord function as
\begin{equation}
\label{eq6.12}
\left| A_j(\xi ) \right|^{-2}  = 
\left(\dot{x}_I \right)_+ \wedge 
\left(\dot{x}_I \right)_- = \left\{ I_+,I_- \right\} \; ,
\end{equation}
where the displaced actions are defined by
\begin{equation}
\label{eq6.13}
I_\pm(X) = I \left( X \pm \frac{\xi}{2} \right) \; 
\end{equation}
(see Fig.~\ref{fig2}).
In the general case each element of the determinant is such a Poisson
bracket for the different  L action variables \cite{ozorio82}.

For a maximal chord, termed a diameter, such that 
$\dot{x}_-$
becomes parallel to 
$\dot{x}_+$,
the amplitude diverges to generate a caustic of the chord function.
Beyond this boundary $\chi(\xi)$ becomes negligible.
In the limit of small chords, $\dot{x}_{\pm}$ are
almost parallel, and the amplitude may be arbitrarily large. 
In this region the above semiclassical theory ceases to operate and, 
because the chord normalization condition is precisely 
$\chi(0)=1/2\pi \hbar$, we have neglected to define the overall 
normalization factor. 
Actually this limit around the origin is much more nasty than normal
semiclassical caustics, since the entire classical manifold can be
defined by a succession of infinitesimal chords. 
It is thus a nontrivial problem to construct a semiclassical uniform
theory including large and small chords.

However, it is possible to connect the semiclassical theory to that 
for the small chords of the previous section. 
Let us take the case that the given chord, $\xi$, is parallel to the 
$q$ axis, so that we can rewrite (\ref{eq6.4}) as
\begin{equation}
\label{eq6.14}
\chi(\xi ) = 
\frac{c^2}{(2\pi\hbar)^L}
\int dQ \exp \left\{ \frac{i}{\hbar} 
        \left[ S \left( Q + \frac{\xi_q}{2}, I \right)-
               S \left( Q - \frac{\xi_q}{2}, I \right) 
        \right]
             \right\}
\left| \det \frac{\partial \theta}{\partial Q} \right|_+^{1/2} 
\left| \det \frac{\partial \theta}{\partial Q} \right|_-^{1/2} \; ,
\end{equation}
using (\ref{eq6.2}). 
Expanding
\begin{equation}
\label{eq6.15}
\theta \left(q_{\pm}\right) = \theta (Q) \pm 
\frac{\partial \theta}{\partial Q} \frac{\xi}{2} + \cdots \; ,
\end{equation}
we obtain 
\begin{equation}
\label{eq6.16}
\left| \det  \frac{\partial \theta}{\partial Q} \right|_-^{1/2}
\left| \det  \frac{\partial \theta}{\partial Q} \right|_+^{1/2} =
\left| \det  \frac{\partial \theta}{\partial Q} \right| 
\left[ 1 + \mathcal{O}(\xi^2) \right] 
\end{equation}
whereas
\begin{equation}
\label{eq6.17}
S \left( Q + \frac{\xi_q}{2} , I \right) -
S \left( Q - \frac{\xi_q}{2} , I \right) =
p_I(Q)\, \xi_q + \mathcal{O} \left( \xi^3_q \right) \; .
\end{equation}
Hence, we can cancel the amplitudes in (\ref{eq6.14}) by changing 
the integration variable, $Q \to \theta$:
\begin{equation}
\label{eq6.18}
\chi(\xi ) \simeq  
\frac{1}{(2\pi\hbar)^L} 
\int^{2\pi}_0  \frac{d\theta}{2\pi} \,
e^{i X(\theta )\wedge \xi/\hbar} \; .
\end{equation}
The general form of (\ref{eq6.18}) holds for any choice of origin and
direction of the small chord $\xi$. 

We have here rederived the small chord expression for the chord
function used in the last section in a way that has several
advantages. 
First we notice that Berry's original derivation of the
$\delta$-function approximation for the Wigner function 
also linearized the action for small chords \cite{berry77}. 
But here the chord is indeed fixed and can be assumed
to be small, whereas there the integral is over all chords. 
We should note that (\ref{eq5.3}), or the extension of (\ref{eq6.18})
for all chords, is tantamount to approximating 
\begin{equation}
\label{eq6.185}
\hat{\rho} \approx 2^L  
\int^{2\pi}_0  \frac{d\theta}{(2\pi)^L} \,
\hat{R}_{X(\theta)} \; .
\end{equation}

\begin{figure}
\includegraphics[height=10cm]{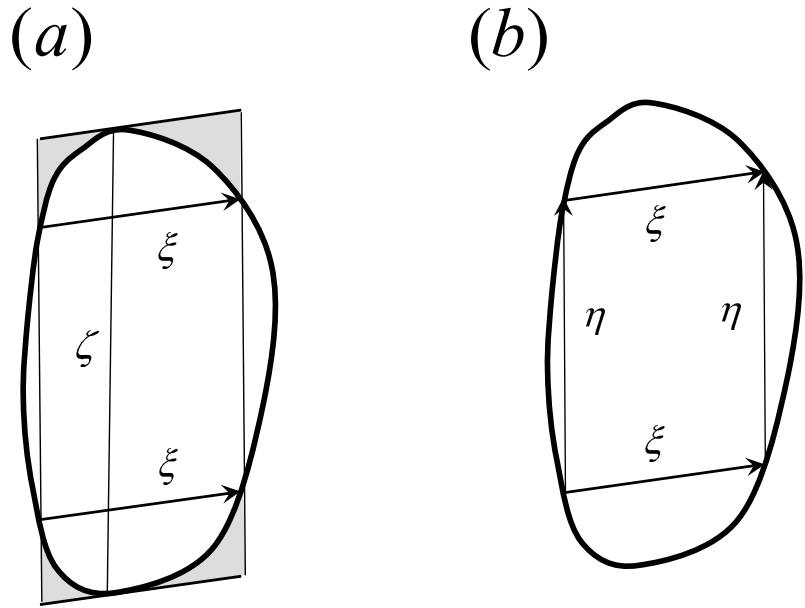}
\caption{(a) The difference in the pair of stationary phases for 
(\ref{eq6.18}) is proportional to the parallelogram that has $\xi$
as one of its sides and is doubly tangent to the quantized curve.
The difference between this area and the pair of stationary
areas in the full semiclassical formula is just the shaded area
at the corners. 
(b) A pair of stationary phase realizations in a convex 
quantized curve for a finite chord $\xi$ defines an inscribed 
parallelogram.
The other side is the conjugate chord $\eta$.
The sign of the conjugate chords plays no role in the equations
for $|\chi(\xi)|^2$ because $\chi(\xi)^\ast=\chi(-\xi)$.}
\label{fig3}
\end{figure}

We can now evaluate the error in (\ref{eq6.18}) when the chord $\xi$ is
large enough for both this integral and (\ref{eq6.4}) to be evaluated
by stationary phase. 
Consider the convex curve $I(x)=\mathcal{I}$ in 
Fig.~\ref{fig3}(a).
The difference in the phases for the stationary phase evaluation at
each stationary point of (\ref{eq6.18}) or the full integral 
(\ref{eq6.4}) is the part of the area near the corner of the
circunscribed parallelogram with two sides $\xi$ tangent to the curve,
lying outside the curve. 
If we approximate the curve as a parabola around the point of 
tangency $\delta p = a (\delta q)^2$, the leftover area is just 
$A(\xi )= 2a|\xi |^3/3$.
So, if the diameter between the parallel tangents with the 
direction $\xi$ is $\zeta(\xi )$, the true stationary area for each 
realization of the chord is just
\begin{equation}
\label{eq6.19}
S(\xi ) = \frac{1}{2} \, \zeta \wedge \xi -\frac{2a|\xi |^3}{3} \; .
\end{equation}
Thus the stationary phase evaluation of the full integral is valid as
long as $|\zeta \wedge \xi| > \hbar$, whereas the integral (\ref{eq6.18}) can 
be used if $a|\xi |^3 < \hbar$. 
Therefore the range of overlap for both approximations is
\begin{equation}
\label{eq6.20}
\frac{\hbar}{|\zeta|} \lesssim  |\xi | \lesssim 
\left( \frac{\hbar}{a} \right)^{1/3} \; ,
\end{equation}
so that we can always make the transition between both approximation
as $\hbar \rightarrow 0$.
%

\section{Conjugate Chords}
\label{section7}

The pair of parallel tangents shown in Fig.3(a) can be
defined as the realizations of an infinitesimal chord for the
curve $I(x)=\mathcal{I}$. The points where the tangents touch
this curve are joined by the diameter $\zeta$.
In the case of a finite chord $\xi$, we can also define the 
conjugate chord, $\eta$, as that which closes off the inscribed 
parallelogram with the sides $\xi$, as shown in 
Fig.~\ref{fig3}(b).
Clearly, the area of this parallelogram is $\xi \wedge \eta$, 
which is smaller than $\xi \wedge \zeta$, used in the previous section. 
Choosing the origin at the centre of this parallelogram, we
obtain the phase of each contribution to $\chi(\xi )$ as 
$\xi \wedge \eta /2$, 
added to each of the chord areas for $\xi$. 

Increasing the length of $\xi$ while keeping its direction constant,
the pair of realizations eventually coalesce along a diameter of the
closed curve. 
Thus we obtain in this limit an interchange in the role of the 
``conjugate chords'' $\xi$ and $\eta$. 
Now it is the small chord $\eta$ that determines the phase of the 
contribution as the chord $\xi$ approaches the caustic of $\chi(\xi )$ 
at its diameter.

\begin{figure}
\includegraphics[width=17cm]{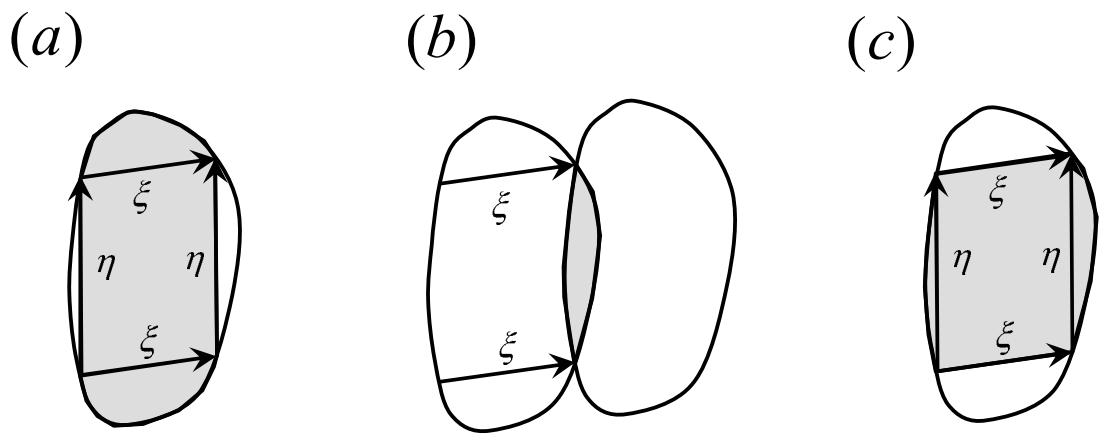}
\caption{The phase difference for the pair of realizations of the chord
$\xi$ is the shaded area in (a) divided by $\hbar$.
The quantization of the curve equates this phase (modulo $2\pi$)
to the phase for the shaded area in (b), obtained by the overlap
of the curve with its translation by $\xi$.
Adding to this the area of the inscribed parallelogram, we obtain
the phase difference for the pair of realizations of the chord 
$\eta$ in (c).}
\label{fig4}
\end{figure}

It is not only in the geometry of the curve that the conjugate
chords $\xi$ and $\eta$ interact. 
Indeed, the phase difference between both realizations of the chord 
$\xi$ coincides with the sum of the chord areas for the realizations 
of the chord $\eta$. 
To see this, notice that this phase difference is just the shaded area 
in Fig.~\ref{fig4}(a) divided by Planck's constant. 
But, because this curve is quantized (its enclosed area is an integer 
factor of $2\pi \hbar$, plus a Maslov correction), the phase difference 
for $\xi$ is the same as for the unshaded area.
Notice now that the phase difference can also be obtained by
translating the whole quantized curve by the chord $\xi$ and
measuring the overlap. 
This is just the classical correspondence for the definition of 
$\chi(\xi )$ of a pure state in (\ref{eq4.7}). 
If the classical quantized curves all have a symmetry centre, 
it makes no difference whether or not the curve is reflected
about its centre prior to translation. 
In this way we obtain a semiclassical version of the trivial 
relation between the Wigner function and the chord function in 
the case of centre-symmetrical systems. 
The overlap between the quantized curves, or quantized tori,
was already employed to study the semiclassical approximation to the
Wigner function \cite{ozorio82}. 

Both the Wigner functions and the chord function can be considered as 
special cases of the overlap of two semiclassical states 
$|\psi \rangle$ and $|\phi \rangle$, corresponding to
classical curves, or Lagrangian surfaces. 
In the case of the Wigner function 
$|\phi \rangle = \hat{R}_X |\psi \rangle$, 
whereas, for the chord function 
$|\phi \rangle = \hat{T}_{\xi} |\psi \rangle$. 
In the general formalism of Littlejohn \cite{littlejohn95}, each 
intersection of the manifolds corresponding to
$|\psi \rangle$ and $|\phi \rangle$ 
determines a semiclassical contribution. 
The phase difference between these contributions is determinated 
by the overlap area, just as in Fig.~\ref{fig4}(b), within a Maslov 
correction. 
The amplitude of each semiclassical term is then given by Poisson
brackets between both manifolds at the intersections. 
Obviously, in the present case, this coincides with the Poisson 
brackets for the same manifold at either tip of the chord, as obtained 
in (\ref{eq6.12}). 
Perharps this general point of view provides the more pleasing 
explanation for the identical form of the Wigner and the chord 
amplitudes.

The relation between conjugate chords, $\xi$ and $\eta$, and the
corresponding evaluations of the semiclassical chord function for the
same state $|\psi \rangle$, i.e., 
$\chi_{\psi}(\xi)$ and $\chi_{\psi}(\eta)$, 
is mediated by the parallelogram formed by $\xi$ and $\eta$. 
The phase difference $\Delta S(\xi )$, in units of $\hbar$,
between both contributions to $\chi_{\psi}(\xi )$ can be pictured as
either the shaded area in Fig.~\ref{fig4}(a), or the unshaded area 
corresponding to both realization of the chord $\eta$. 
But, if we now add $\xi \wedge \eta$ to these unshaded areas, we obtain
the new shaded area $\Delta S(\eta)$ in Fig.~\ref{fig4}(c), which 
determines the phase difference of both contributions to 
$\chi_{\psi} (\eta )$. 

Squaring the semiclassical approximation for $\chi(\eta)$
we obtain 
\begin{equation}
\label{eq7.1}
(2\pi\hbar)^2 C_{\eta} = |\chi(\eta )|^2 \simeq 
\{I_+,I_-\}^{-1}_1 +
\{I_+,I_-\}^{-1}_2 +
\{I_+,I_-\}^{-\frac{1}{2}}_1
\{I_+,I_-\}^{-\frac{1}{2}}_2  \cos \frac{\Delta S(\eta)}{\hbar} \; 
\end{equation}
in the simple case where $I(x)=\mathcal{I}$ is a convex curve, so
that there is a pair of realizations for the chord, $\eta$, 
with intensities given by (\ref{eq6.12}).
Inserting (\ref{eq7.1}) this into the Fourier integral 
(\ref{eq4.8}), we find that 
the first two smooth terms only contribute to the classical 
neighbourhood of the origin.
It turns out that the stationary condition for $\eta$ in (\ref{eq4.8}) 
for a given value of $\xi$ in the last term is precisely that $\eta$ 
be the conjugate chord to $\xi$, i.e., that $\xi$ and $\eta$ form an 
inscribed parallelogram in the curve corresponding to $|\psi \rangle$. 
This is a consequence of equation (\ref{eq6.5}). 
Therefore the Fourier invariance of the quantum correlation results 
semiclassically from the relation of the chord function 
$\chi(\xi )\leftrightarrow \chi(\eta)$, for conjugate chords.

In the limit as the family of chords with a fixed direction
approaches its maximum value for a given quantized curve, i.e., a
diameter, the conjugate chord approaches the origin in the direction
of the parallel tangents at the tips of the diameter. 
However, in this region, the simple semiclassical amplitudes $A_j$ 
given by (\ref{eq6.12}) become singular and each of the three terms 
in (\ref{eq7.1}) contribute. 
It is here necessary to replace the semiclassical contribution by 
uniform approximations in terms of Airy functions. This will be
the subject of further work.

The role of diameters as conjugate chords near the origin is 
brought forth by combining (\ref{eq4.8}) with (\ref{eq5.5}), 
for $|\xi| \to 0$:
\begin{equation}
\label{eq7.2}
C_{\xi} \approx
\int 
\frac{d\theta_+}{(2\pi)^L} \,
\frac{d\theta_-}{(2\pi)^L} \,
e^{i \xi \wedge \left[ x (\theta_+) 
                    -x (\theta_-) \right]/\hbar} \; .
\end{equation}
We can interpret $x(\theta_+)-x(\theta_-)$ as the set of chords
supported by the quantized curve, or $L$-dimensional torus. 
This confirms the role of the first two terms of (\ref{eq7.1}) which 
contribute to the Fourier transform of $C_{\xi}$ for small $\xi$: 
all the chords supported by the curve contribute. 
However, as soon as $\xi$ grows in modulus enough that we may evaluate  
(\ref{eq7.2}) by stationary phase [condition (\ref{eq6.20})], the 
dominant contribution comes from the diameter $\zeta (\xi )$, i.e.,
$\zeta = x(\theta_+)-x(\theta_-)$,
such that 
\begin{equation}
\label{eq7.3}
\frac{dx}{d\theta_+} = \frac{dx}{d\theta_-} = 0 \; ;
\end{equation}
in other words, the tangents at $\theta_{\pm}$ are parallel.

Up to now we have analyzed cases where the classical structure 
is a continuous curve or surface in phase space.
Let us now consider an alternative classical setting, a state
\begin{equation}
\label{eq7.4}
|\psi \rangle = \sum_j a_j |\eta_j \rangle \; ,
\end{equation}
where $|\eta_j \rangle$ are coherent states centered on the phase 
space points $\eta_j$. 
Then the density operator is
\begin{equation}
\label{eq7.5}
|\psi \rangle \langle \psi | = 
\sum_j |a_j|^2 |\eta_j \rangle \langle \eta_j| +
\sum_{j \ne k} a_j a^\ast_k |\eta_j \rangle \langle \eta_k| \; ,
\end{equation}
and its chord representation $\chi_{\psi}(\xi )$ is a simple generalization
of (\ref{eq4.3}): each chord $\pm (\eta_j-\eta_k)$ is the centre
of a gaussian, whereas the diagonal terms in (\ref{eq7.5}) interfere
collectively in the neighbourhood of the origin. 
Taking 
$\left| \chi_{\psi}(\xi ) \right|^2$, 
we again have gaussians centred at $\eta_j-\eta_k$ and at the origin. 
It is easy to see that the Fourier transform merely interchanges 
the contributions to the origin with the pair of gaussians centred at 
$\pm (\eta_j-\eta_k)$. 
However, there is a new contribution to the Fourier transform if the
four vectors 
$\eta_1, \eta_2, \eta_3, \eta_4$ 
form a parallelogram, i.e.,
$\eta_2 - \eta_1 \simeq \eta_3 - \eta_4$ 
and consequently
$\eta_3 - \eta_2 \simeq \eta_4 - \eta_1$, 
where $\simeq$ means that the vectors differ by 
$\mathcal{O}(\sqrt{\hbar})$.
Then the Fourier transform of $|\chi(\xi )|^2$ for 
$\xi =\eta_1-\eta_4$ 
receives contributions from the neighbourhood 
of $\eta_2-\eta_1$ 
and vice-versa. 
Thus, again we verify the role of conjugate chords in
the Fourier invariance of the quantum correlations. 
The consistency of fitting semiclassical states 
with gaussian coherent states has been recently investigated
in \cite{kenfack04}.

\section{Discussion: Resurgence of pure state correlations}
\label{section8}

We recapitulate our findings: As we displace a pure state that is classically
extended, i.e. that quantizes with a large quantum number, the correlation
in a given direction goes through four different stages: a) An initial, relatively
simple stage of very short chords in which the behaviour is quadratic in the
displacement and determined purely by the phase space extent of the state.
This is the region that has received most attention in the litterature so far 
\cite{zurek01,jordan01,alonso04}.
b)A second oscillatory stage ruled by the points of intersection of the two
displaced tori, with relative phases that are semiclassically determined.
c) A third stage characterized by a chord caustic where the semiclassical
contributions diverge and where uniform approximations are still needed.   
d) An asymptotic region where the correlation decays uniformly to zero.
We have here given special attention to the transitions between stages (a) and (b) . 
Other large systems such as a widespread superposition of coherent states 
will also have large classical values for $\langle p^2\rangle$ or $\langle q^2\rangle$. 
For all such pure states the initial decay of the correlations 
with growing displacement will be followed by their oscillatory resurgence 
for those chords that cause the underlying classical structures to overlap. 
 
It is important to consider higher phase space dimensions, 
since all our examples were restricted to a single degree of freedom. 
Again, it is clear that in the case of arbitrary superpositions of 
pairs of coherent states the straight Fourier analysis of the Wigner 
function leads to very similar pictures for the pair of classical-like 
gaussians in the Wigner function and the peaks of correlations at 
the chords that separate them.

Integrable systems of higher dimension might appear to be harder to
analyze, but this is not so. 
It was show in Ref.~\cite{ozorio82} that the Wigner function corresponding 
to $L$-dimensional quantized tori are characterized by a Wigner caustic 
of dimension $2L-1$ in which the torus appears as a higher singularity. 
At this boundary between an oscillatory inner region of phase space and 
the evanescent region outside, the Wigner function attains an amplitude 
maximum, which can be described locally by an Airy funtion and its 
derivative. 
In the case that there exists a reflection symmetry centre, the chord
function  must be identical to the Wigner function within a phase and 
the rescaling  (\ref{eq3.14}). 
Even  without such a reflection symmetry, the region 
where a finite $L$-torus and its rigid translation intersect is bounded 
by a $(2L-1)$-dimensional surface at which they touch nontransversally 
-- the chord caustic. 

Finally we must consider the case of an eigenstate of a chaotic
Hamiltonian. 
In this case Shnirelman's theorem \cite{shnirelman}
guarantees that most states are 
ergodic in the sense that the average of smooth functions
of the observables $\hat{p}$ and $\hat{q}$ are given by a
classical average over the energy shell. 
In its simplest form, quantum ergodicity may be taken as the
Berry-Voros hypothesis that the Wigner function is approximately
a Dirac delta function on the energy shell \cite{wfc, voros76}.
It follows immediately that for small chords we may adapt the discussion 
in section V to obtain the chord function for the n'th energy eigenstate
as
\begin{equation}
\label{eq8.1}
\chi_n (\xi ) = \langle e^{i \xi \wedge \hat{x}} \rangle_n =
\frac{1}{(2\pi \hbar)^L}
\frac{ \int dx \, \delta \left( H(x)-E_n \right) \ e^{i \xi \wedge x/\hbar}}
     { \int dx \, \delta \left( H(x)-E_n \right)} 
\end{equation}
(see also \cite{jordan01,alonso04}).
Conjugate to this short chord behaviour there must be appreciable large
arguments for which $\chi_n(\xi )$ is also large, corresponding to the largest
chord fitting between a pair of points in the energy shell for any
given phase space direction. 
Once again, in the case of a reflection symmetric chaotic Hamiltonian, we may 
invoke the (scaled) identity with the Wigner function which is dominated 
by the energy shell itself, according to Shnirelman's theorem 
\cite{shnirelman}.

In all the extensions to higher dimensions, large chords $\xi$, generate
oscillations in the Wigner function with wave vector $J\xi$.
In the case of a semiclassical state constructed on a Lagrangian torus,
the fine structure in the Wigner function results from the interference of 
a finite number of locally plane waves.
The fine structure of ergodic chaotic Wigner functions has not yet been
explored. 

It should be noted that our general description is in no way limited to 
stationary states. The curves and surfaces which we have mainly treated
may be evolving classically, while the corresponding quantum system also evolves.
We have not treated here this dynamics, but our description is valid for any 
snapshot. Even for the evolution of a localized wave packet in a chaotic system, 
it is possible to advance that long scale quantum correlations develop 
as the classical packet spreads over the energy shell.  

This study of the coherence properties of pure states
suggests that it is not always profitable to transform back to 
the centre phase space of the Wigner function from the phase
space of chords, since it is here that the full structure of 
chord conjugacies is manifest.
Furthermore, there is a decided advantage to allow the collapse
of all classical information onto the neighborhood of $\xi=0$.
All that extends out in the chord phase space are signs of 
quantum coherence and these are structures that are lost
in the nonunitary evolution that results from tracing out the
interaction with an uncontrollable environment.

\begin{acknowledgments}
We thank O. Brodier for interesting comments.
Partial financial support from 
Millenium Institute of Quantum Information, PROSUL, CAPG-BA (CAPES) 
and CNPq is gratefully acknowledged. 
\end{acknowledgments}

\appendix

\section{Conjugate Phase Spaces}
\label{appA}

The points in phase space for a system with $L$ degrees of freedom
are here denoted simply by 
$x=(p,q)=\left(p_1, \cdots , p_L,q_1, \cdots , q_L\right)$. 
In the classical limit, $\hbar \rightarrow 0$, we may
associate a quantum state to such a phase  point, but we need pairs
of points $(x_-,x_+)$ to describe operators $\hat{A}$, such as 
$\langle q_+|\hat{A}|q_-\rangle$, 
or 
$\langle p_+|\hat{A}|p_-\rangle$. 
In both these cases, we only use, in fact, half of the phase space 
variables, as consistent with the uncertainty principle.
 
Another alternative is to use either the {\em centre}
\begin{equation}
\label{eqc.1}
X=(P,Q)=(x_++x_-)/2
\end{equation}
or the {\em chord}
\begin{equation}
\label{eqc.2}
\xi = \left( \xi_p,\xi_q \right) = x_+ - x_-  \; .
\end{equation}
As shown in \cite{berry77,ozorio98} we may then identify the argument in the 
Weyl symbol of $\hat{A}$, $A(X)$, with the center (\ref{eqc.1}).
In the same way, the argument in the chord symbol $A(\xi )$ 
corresponds to the chord (\ref{eqc.2}). 
In the case of a unitary transformation, corresponding classically 
to a trajectory, we identify $\xi$ as the chord corresponding 
to the arc
$\{x(\tau), \; 0\le \tau \le t \}$ 
joining $x_-=x(0)$ and $x_+=x(t)$. 
Of course, $X$ is the reflection centre for this pair of phase space 
points \cite{ozorio98}. 

Canonical transformations in phase space are generated implicitly by
the centre action, i.e., the generating function 
$S(X)$ \cite{ozorio98} such that
\begin{equation}
\label{eqc.3}
J\xi = \frac{\partial S}{\partial X} \; ,
\end{equation}
where
\begin{equation}
\label{eqc.4}
J=\left(
\begin{array}{cc}
    0    & -\openone \\ 
\openone &      0 
\end{array}
\right) \; , 
\end{equation}
or by the chord action $S(\xi )$, such that
\begin{equation}
\label{eqc.5}
-JX=\frac{\partial S}{\partial \xi} \; .
\end{equation}
Comparing with more familiar generating functions, e.g., $S(q_-,q_+)$,
\begin{equation}
\label{eqc.6}
\frac{\partial S}{\partial q_+}=p_+\ , \qquad
\frac{\partial S}{\partial q_-}=p_- \ ,
\end{equation}
it follows that the conjugate variable to $X$ should really be 
$J \xi$ rather than the chord itself. However, this lacks a clear
geometrical interpretation in terms of $x_\pm$, so it is better to 
keep $\xi$, but replace all scalar products by skew products. 
Hence, in all Fourier transforms we use
\begin{equation}
\label{eqc.7}
(J\xi )\cdot X = \xi \wedge X = \xi_p \cdot Q - \xi_q \cdot P =
\sum^L_{\ell =1}
\left(\xi_{p_{\ell}} Q_{\ell} - \xi_{q_{\ell}} P_{\ell} \right) \; .
\end{equation}

\section{Parity Eigenstates}
\label{appB}

Any state $|\psi \rangle$ in Hilbert space can be decomposed into
components of even or odd parity (eigenvalue $+1$, or $-1$) of any
of the reflection operators $\hat{R}_X$. 
Indeed these projectors were presented by Royer \cite{royer77} as
\begin{equation}
\label{eqp.1}
\hat{P}^X_\pm = \frac{1}{2}
\left( 1 \pm \pi \hbar \hat{R}_X \right) \; ,
\end{equation}
so that the parity decomposition of an arbitrary density operator,
$\hat{\rho}$, is
\begin{equation}
\label{eqp.2}
\hat{\rho}^X_{\pm} = 
\frac{\hat{P}^X_{\pm} \ \hat{\rho}\ \hat{P}^X_{\pm}}
{   \tr   \ \hat{\rho} \ \hat{P}^X_{\pm}} \; .
\end{equation}
Since these reduced density operators commute with the reflection
operator, the Wigner function corresponding to $\hat{R}_X \hat{\rho}$ 
is
\begin{equation}
\label{eqp.3}
W_X^{\pm}(X') = 2^L \tr \hat{R}_{X'} \hat{R}_X \hat{\rho}^X_{\pm} =
\pm \ 2^L \tr \hat{R}_{X'} \hat{\rho}^X_{\pm} = 
      2^L e^{ 2i \ X' \wedge X / \hbar } \tr \hat{T}_{2(X'-X)} 
\hat{\rho}^X_{\pm} \; ,
\end{equation}
where we used the general group relations between translations and
reflections \cite{ozorio98}. 
Thus, shifting the origin to the centre of symmetry, we obtain
\begin{equation}
\label{eqp.4}
W^{\pm}_0(X)=\pm \ 2^L\chi^{\pm}_0(-2X) \; .
\end{equation}
In the case that $\hat{\rho}^X_{\pm}$ is the projection of an arbitrary
$\hat{\rho}$ according  to (\ref{eqp.2}), then the general form of the
projected Wigner function given by \cite{ozorio04} is transported to the
chord function  as 
\begin{equation}
\label{eqp.5}
\chi^{\pm}_0(\xi )=\frac{\chi(\xi )+\chi(-\xi )\pm 2W(\xi/2)}
{4\left[1\pm \pi \hbar W(0)\right]} \; .
\end{equation}

Evidently, all chord functions with pure parity are real. The
reciprocal is also true, because the imaginary part of $\chi(\xi )$ 
cancels if and only if
\begin{equation}
\label{eqp.6}
\tr (\hat{T}_{-\xi} \hat{\rho}) = \chi( \xi )= \chi(\xi )^{\ast}=
\chi(-\xi )= \tr  \hat{T}_{\xi} \hat{\rho } \; .
\end{equation}
But, using the group properties of translations and reflections,
\begin{equation}
\label{qp.7}
\tr \hat{T}_{-\xi} \hat{\rho} = 
\tr \hat{R}_0 \hat{T}_{\xi} \hat{R}_0 \hat{\rho} = 
\tr \hat{T}_{\xi} \hat{R}_0 \hat{\rho} \hat{R}_0 \; ,
\end{equation}
which is only equal to (\ref{eqp.6}) for all $\xi$ if 
$[\hat{\rho},\hat{R}_0]=0$. 

A mixture of pure states, each of which has definite parity, commutes
with $\hat{R}_0$ and hence produces a real chord function. 
The imaginary part of the chord function is related to the off-diagonal
parity representation of the density matrix, i.e., for an orthogonal
basis of odd and even states, this is the block of the density matrix
coupling the different parities.

\newpage



\begin{thebibliography}{99}


\bibitem{zurek01}
W. H. Zurek, Nature {\bf 412}, 712 (2001).

\bibitem{peres84}
A. Peres, Phys. Rev. A {\bf 30}, 1610 (1984).

\bibitem{loschmidt}
R. A. Jalabert and H. M. Pastawski, 
Phys. Rev. Lett. {\bf 86}, 2490 (2001); 
Ph. Jaquod, P. G. Silvestrov and C. W. J. Beenakker,
Phys. Rev. E {\bf 64}, 055203 (2001); 
F. M. Cucchietti, H. M. Pastawski, and D. A. Wisniacki, 
Phys. Rev. E {\bf 65}, 045206 (2002); 
G. Benenti and G. Casati, 
Phys. Rev. E {\bf 65}, 066205 (2002);
T. Prosen and M. Znidaric, J. Phys. A {\bf 34}, L681 (2001); 
T. Prosen, Phys. Rev. E {\bf 65}, 036208 (2002).

\bibitem{garciamata04}
I. García-Mata and M. Saraceno, 
Phys. Rev. E {\bf 69}, 056211 (2004).

\bibitem{alonso04}
D. Alonso, S. Brouard, J. P. Palao and R. S. Mayato, 
Phys. Rev. A {\bf 69}, 052111 (2004).
 
\bibitem{wfc}
M. V. Berry,
J. Phys. A. {\bf 10}, 2083 (1977).

\bibitem{schleich}
P. W. Schleich, 
\textsl{Quantum Optics in Phase Space} (Wiley-VCH, Berlin, 2001).

\bibitem{glauber63}
R. J. Glauber, 
Phys. Rev. {\bf 131}, 2766 (1963).

\bibitem{lutterbach97}
L. G. Lutterbach and L. Davidovich,
Phys. Rev. Lett. {\bf 78}, 2547 (1997).

\bibitem{caves04}
C. M. Caves and K. W\'odkiewicz, 
e-print quant-ph/0409063.

\bibitem{royer77} 
A. Royer, 
Phys. Rev. A {\bf 15}, 449 (1977).

\bibitem{ozorio98} 
A. M. Ozorio de Almeida,
Phys. Rep. {\bf 295}, 265 (1998).

\bibitem{coxeter} 
H. S. M. Coxeter,
\textsl{Introduction to Geometry} (Wiley, New York, 1961).

\bibitem{brodier04}
O. Brodier and A. M. Ozorio de Almeida,
Phys. Rev. E {\bf 69}, 016204 (2004).

\bibitem{gronewold46}
H. J. Gr\"onewold,
Physica {\bf 12}, 405 (1946). 

\bibitem{berry77}
M. V. Berry, 
Phil. Trans. R. Soc. London {\bf 287}, 30 (1977).

\bibitem{abramowicz}
M. Abramowitz and I. A. Stegun,
\textsl{Handbook of Mathematical Functions} 
(Dover Publications, New York, 1964).

\bibitem{vanvleck28}
J. H. Van Vleck,
Proc. Natl. Acad. Sci. USA {\bf 14}, 178 (1928).

\bibitem{ozorio}
A. M. Ozorio de Almeida,
\textsl{Hamiltonian Systems: Chaos and Quantization}
(Cambridge University Press, Cambridge, 1988).

\bibitem{ozorio82}
A. M. Ozorio de Almeida and J. Hannay, 
Ann. Phys. {\bf 138}, 115 (1982).

\bibitem{littlejohn95}
R. G. Littlejohn, in 
\textsl{Quantum Chaos: Between Order and Disorder},
edited by G. Casati and B. Chirikov, 343-404 
(Cambridge University Press, Cambridge, 1995).

\bibitem{kenfack04}
A. Kenfack, J. M. Rost and A. M. Ozorio de Almeida,
J. Phys. B {\bf 37}, 1645 (2004).

\bibitem{jordan01}
A. Jordan and M. Srednicki, e-print quant-ph/0112139.

\bibitem{shnirelman}
A. I. Shnirelman,
Uspehi. Mat. Nauk. {\bf 29}, 181 (1974);
Y. Colin de Verdi\`ere, 
Comm. Math. Phys. {\bf 102}, 497 (1985);
S. Zelditch, 
Duke Math. J. {\bf 55}, 919 (1987).

\bibitem{voros76}
A. Voros, 
Ann. Inst. Henri Poincaré {\bf 24A}, 31 (1976).


\bibitem{ozorio04}
A. M. Ozorio de Almeida and O. Brodier,
J. Phys. A {\bf 37}, L249 (2004).

\end{thebibliography}
\end{document}